\documentclass[12pt]{article}
\usepackage{graphicx}
\usepackage{amssymb}
\usepackage{xcolor}
\usepackage{cite}
\usepackage[utf8]{inputenc}

\begin{document}

\title{Solar Neutrino Limits on Decoherence}

\author{P. C. de Holanda}

\maketitle

\begin{abstract}
  The solar neutrino flux arrives at Earth as an incoherent admixture of mass eigenstates, and then solar neutrino detection constitute a blind probe to the oscillation pattern of the neutrino flavour conversion. Consequently, it is also impossible to probe, in a model independent approach, any new physics that leads to an enhancement of decoherence during the neutrino evolution, an effect that is present for instance in Open Quantum System formalism. However, such mechanism can also induce changes between mass eigenstates if an energy interchange between the neutrino subsystem and the reservoir is not explicitly forbiden. In this work the conversion probabilities between mass eigenstates in an Open Quantum System are calculated, and limits are stablished for these kind of transitions. We present our results in a pedagogical way, pointing out how far the analysis can go without any assumption on the neutrino conversion physics inside the Sun, before performing the full calculations. We obtain as limits for the decoherence parameters the values of $  \Gamma_3< 6.5\times 10^{-19}$ eV and $\Gamma_8< 7.1\times 10^{-19}$ eV at 3$\sigma$.
\end{abstract}

\section{Introduction}
The supression in the total flux of solar neutrinos in comparison with
the theorethical predictions provided the first indication of
neutrino flavour conversion. The first results of the Homestake
experiment already indicated that the observable flux of
electron neutrinos arriving from the Sun were in clear disagreement with the
the solar models predictions at the time~\cite{Bahcall:1964gx,Bahcall:1976zz},
a result that was later confirmed by a number
of other experiments. Sage~\cite{Abazov:1991rx} in 1991 and Gallex~\cite{Anselmann:1992kc} in 1992, Kamiokande~\cite{Hirata:1989zj} in 1988 and Super-Kamiokande~\cite{Fukuda:1998fd} in 1998 and
SNO~\cite{Ahmad:2001an} in 2001, all confirmed the suppression on the electron neutrino flux in
different parts of the solar neutrino spectrum, while confirming the main
features of the Standard Solar Model.%~\cite{SSM} (SSM).

The flavour conversion induced by a non-vanishing neutrino mass and a mixing matrix relating flavour and mass eigenbasis was the first proposed solution to such discrepancy~\cite{Pontecorvo:1967fh}. It soon became clear that neutrino interactions with solar matter would have a fundamental role on its flavour conversion through resonant amplification, the so-called  MSW effect~\cite{PhysRevD.17.2369,Mikheev:1986gs,Mikheyev1986}. But several other mechanisms could also provide flavour conversion compatible with the data, and by the end of the 90's there were a number of possible solutions (see, for instance~\cite{Gago:2001si} for a comprehensive comparison between possible explanations for the solar neutrino problem).

Only in 2002 the results from KamLAND~\cite{Eguchi:2002dm}, a reactor neutrino detector, confirmed the mass induced flavour conversion of solar neutrinos. Located at Kamioka mine, KamLAND detected electron anti-neutrinos created in several different nuclear reactors on Japan's territory. Its results were compatible with flavour neutrino oscillations with oscillation parameters compatible with the solution to the solar neutrino problem based on resonant conversion with a large mixing angle - LMA. Several recent results confirm such scenario with great precision (see for instance~\cite{deSalas:2017kay} and~\cite{Esteban2019} for comprehensive oscillation results analyses).

As usually happens with scientific explanations, the correct chronology of the
events are not always the most pedagogical one. In nowadays is more
straightforward to invert the chronology and present the oscillation phenomena
starting with the vacuum oscillations observed by accelerator and reactor experiments, such as KamLAND,  and using the solar neutrinos as an atemporal comprovation of KamLAND's result.

In accordance with this pedagogical strategy, in this paper I present an exercise on how much information we can extract on solar neutrino flux independent of the flavour conversion mechanisms that operates inside the Sun. Departing from the safe assumption that the solar neutrino flux arrives at Earth as an incoherent admixture of mass eigenstates, I analyze what can be inferred about the fraction of each eigenstate in the total flux for different neutrino energies, using only information from the allowed regions for neutrino oscillation parameters provided by terrestrial experiments such as KamLAND~\cite{Decowski:2016axc} for the neutrino parameters $\theta_{12}$ and $\Delta m^2_{21}$ and Daya-Bay~\cite{Adey:2018zwh}, Reno~\cite{Bak:2018ydk} and Double-Chooz~\cite{Abe:2014bwa} for the neutrino parameter $\theta_{13}$, besides the predictions for the total solar neutrino flux and spectrum~\cite{Bahcall:2004pz,Vinyoles:2016djt}.

The knowledge of the solar neutrino flavour conversion mechanism inside the Sun clearly allows a more comprehensive analysis, but this may not be the case with other astrophysical environments that produces neutrinos. We then use the exercise presented here to extrapolate how other neutrino astrophysical sources could be analysed.

In section~\ref{sec:nuflux} we infer by present experimental data on solar neutrinos what information can be extracted about the solar neutrino flux partition between mass eigenstates, for low energy and high energy neutrinos. In section~\ref{sec:decoprob} we compute the effect of Open Quantum System formalism on neutrino conversion probabilities. In section~\ref{sec:results} I obtain a limit on relaxation effects induced by open quantum system formalism, both independent from any consideration about the neutrino physics operating inside the Sun, and assuming the MSW-LMA conversion. Finally, I propose that similar procedures can be used to analyse astrophysical neutrinos, where the uncertainties on the flavour conversion mechanisms inside the neutrino astrophysical source are considerably greater than for solar neutrinos. In section~\ref{sec:conclusions} I draw the conclusions.

\section{Solar Neutrino's Flux Constraints}\label{sec:nuflux}

The solar neutrino flux arrives at the Earth detectors as a complete
incoherent admixture of mass eigenstates. Any coherence is lost due to
three effects that solar neutrinos are subjected~\cite{deHolanda:2003tx}:
\begin{itemize}
  \item the average over the
    neutrino production point inside the sun;
  \item the fast oscillation on energy;
  \item the wave package decoherence induced by neutrino evolution.
\end{itemize}
Then, due to decoherence, the
solar neutrino flavour conversion probabilities can be written as:
\begin{equation}
P_{e\beta}=\sum_i P_{e i}^{Sun} P_{i\beta}^{Earth}
\label{eq:peipie}
\end{equation}
where $P_{ei}^{Sun}$ is the probability that an electron neutrino produced in the
sun interior arrives at the Earth as eigenstate $i$,
while $P_{i\beta}^{Earth}$ is the probability of detecting a neutrino that arrives
at Earth as eigenstate $i$ as flavour $\beta$. In this section we extract the possible values for $P_{e i}^{Sun}$ from the available solar neutrino data.

\subsection{Neutrino's flavour survival probabilities}

As it has been done in qualitative analysis regarding solar neutrinos, it
is convenient to analyse the neutrino probabilities in three different energy
ranges, low, medium and high energy neutrinos.

\subsubsection{high energy:} \label{sec:highenergy}
The high energy neutrino flux consists mainly of Boron neutrinos, and the flux can be determined by experiments Super-Kamiokande~\cite{Hosaka:2005um,Fukuda:2002pe,Abe:2016nxk} and SNO~\cite{Aharmim:2011vm}. Under the assumption that the high-energy neutrino flux ($E\gtrsim 5$ MeV) is well described by a constant
electron neutrino survival probability, the solar data
allows to stablish a value for such probability fiting the data from
the different Super-Kamiokande and SNO phases. Using only the data corresponding to neutrinos arriving at the detectors during the day, the following $\chi^2$ can be used:
\[
\chi^2(f_B,P_{ee})=\chi^2_{SK}(f_B,P_{ee})+\chi^2_{SNO}(f_B,P_{ee})~~~{\rm ,}
\]
where $f_B$ is a free normalization factor for the Boron neutrino flux.

The Super-Kamiokande $\chi^2$ is given by:
\[
\chi^2_{SK}(f_B,P_{ee})=\left[\sum_{ij} (f_BR_{SK}^{th,i}-R_{SK}^{exp,i})
  [\sigma_{SK}]^{-2}(f_BR_{SK}^{th,j}-R_{SK}^{exp,j})\right]
\]
where the sum covers data from all Super-Kamiokande runs, and $\sigma_{SK}$ entails all uncertainties and correlations. By itself Super-Kamiokande daily data can not provide any information on $f_B$ and $P_{ee}$, since there is an almost perfect correlation between both variables.

SNO collaboration already presents their results in terms of survival probabilities, parametrized by 6 parameters: 3 regarding the parameterization of the daily probability $P_{ee}^d$, 2 regarding the regeneration effect $A_{ee}$, and 1 regarding neutral current measurement of total neutrino flux, which can be associated with the free normalization $f_B$ mentioned above:
\begin{eqnarray*}
  P_{ee}^d(E_\nu)&=&c_0+c_1(E_\nu[MeV]-10)+c_2(E_\nu[MeV]-10)^2\\
  A_{ee}(E_\nu)&=&a_0+a_1(E_\nu[MeV]-10)\\
  f_B&=&\phi_B^{SNO}/\phi_B^{SSM}
\end{eqnarray*}
where $\phi_B^{SSM}$ is the Standard Solar Model prediction to the total Boron neutrino flux.

The $\chi^2$ computation is more straightforward, and we have:
\[
\chi^2_{SNO}(f_B,P_{ee})=\left[\sum_{ij} (f_Br_{SNO}^{th,i}-r_{SNO}^{exp,i})
  [\sigma_{SNO}]^{-2}(f_Br_{SNO}^{th,j}-r_{SNO}^{exp,j})\right]
\]
where $r_{SNO}^{exp}=(\phi_B^{SNO},c_0,c_1,c_2)$ and our theoretical input is 
$r_{SNO}^{th}=(f_B\phi_B^{SSM},P_{ee},0,0)$. We disregard the parameters involved in regeneration effect, and then the sum runs in only 4 parameters. $\sigma_{SNO}$ contains the uncertainties and correlations provided by SNO.

With all taking into account, and minimizing over $f_B$, we obtain:
\begin{equation}
P^{HE}_{ee}=0.324\pm 0.014~~~{(\rm 1\sigma)}
\end{equation}

%This value is not to be exactly compared to the fit to parameter $c_0$ published by SNO~\cite{Aharmim:2011vm} (see also table XV at~\cite{Abe:2016nxk}), since it assumes a constant survival probability, excluding the other parameters from the fit.

\subsubsection{intermediate energy:} 
The best choice to stablish a survival probability for neutrinos with intermediate energy is to use data from Borexino~\cite{Agostini:2018uly}. The collaboration reports a survival probability for the  monoenergetic $^7$Be and {\it pep} neutrino fluxes (1$\sigma$):
\begin{equation}
  P_{ee}(^7Be, 0.862\,{\rm MeV})=0.52\pm 0.05~~~;~~~
  P_{ee}(pep, 1.44\,{\rm MeV})=0.43\pm 0.11
\end{equation}

We do not use these values in our analysis, but they are necessary to stablish a probability for the low energy neutrinos, calculated in what follows.

\subsubsection{low energy:}
Using data from Gallex/GNO and Sage it is possible to stablish a value for the survival probability for low-energy neutrinos.
Since Super-Kamiokande and SNO provides a real-time detection, it is possible
to use only data collected during the day, and exclude possible regeneration
effects from our analysis. Low-energy solar neutrino experiments does not allow
such discrimination, so we rely on the assumption that
regeneration effects on Earth does not change the survival probability. This is in accordance to our assumption that any new physics on KamLAND would be subleading, and for the mixing parameters given by KamLAND no Earth regeneration is expected for low energy neutrinos.

Gallex/GNO and Sage detect neutrinos from all the 8 neutrino sources in the Sun. In the absence of flavour conversions, the fraction of each source on the final detection rate is expected to be:
\begin{eqnarray*}
r&=&(0.561,2.34\times 10^{-2},4.5\times 10^{-5},0.279,\\
&&0.104,1.32\times 10^{-2},1.837\times 10^{-2},4.8\times 10^{-4})
\end{eqnarray*}
for neutrinos from the chains (pp, pep, hep, $^7$Be, B, C, N, O), respectively. The three main contributions come from $pp$ neutrino, $^7$Be neutrinos,  $^8$B neutrinos and $pep$ neutrinos, in decreasing order. We can use the range for the probabilities for the three last of these contributions discussed at the last section to extract a probability for $pp$ neutrinos. Performing the following $\chi^2$ analysis:
\begin{eqnarray*}
\chi^2_{LE}(f_B,P_{B},P_{Be},P_{pep})&=&\left[\sum_{ij} (R_i^{th}-R_i^{exp})
  [\sigma_{LE}]^{-2}(R_i^{th}-R_i^{exp})\right]+\\
&&\left(\frac{P_{B}-0.324}{0.014}\right)^2+
\left(\frac{P_{Be}-0.52}{0.05}\right)^2+
\left(\frac{P_{pep}-0.43}{0.11}\right)^2
\end{eqnarray*}
we obtain after marginalizing over the free parameters ($1 \sigma$):
\begin{equation}
P^{LE}_{ee}=0.57\pm 0.06 % 573 634  511 
\end{equation}

Borexino also reports a survival probability for {\it pp} neutrinos fully
consistent with our estimation:
\[
  P^{LE}_{ee}(pp,E<0.42\,{\rm MeV})=0.57\pm 0.09~~~{(\rm 1\sigma)}
\]
leading, in a combined analysis, to a sligthly more restringent value:
\begin{equation}
P^{LE}_{ee}=0.57\pm 0.05 % 573 634  511 
\end{equation}
which we used in what follows.

\subsection{Neutrino's mass eigenstates probabilities}
\label{sec:nuproblimits}
  
For solar neutrinos arriving at the detector during the day (or for low energy neutrinos, in which the Earth regeneration effect can be disregarded), we can directly
replace $P_{i\beta}^{Earth}$ by the corresponding mixing angles:
\begin{eqnarray*}
  P_{ee}&=&P_{e1}^{Sun} |U_{1e}|^2+P_{e2}^{Sun} |U_{2e}|^2+P_{e3}^{Sun} |U_{3e}|^2\\
 &=&c^2_{13}(P_{e1}^{Sun}c^2_{12}+P_{e2}^{Sun}s^2_{12})+s^2_{13}P_{e3}^{Sun} ~~.
\end{eqnarray*}
For neutrinos arriving at the detector during the night, we should
calculate $P_{i\beta}^{Earth}$.

With KamLAND results on $\theta_{12}$ and the recent
measurement of $\theta_{13}$ by Daya-Bay, Double-Chooz and Reno, we can present an
exercise of how much do we know about the physical eigenstates distribution of
solar neutrinos in a complete solar model independent way. 

By defining the following $\chi^2$ related to low and high-energy solar neutrinos
in the following way:
\begin{eqnarray}
  \chi^2_{LE}&=&\left(\frac{P_{ee}(P_{e1}^{Sun},P_{e2}^{Sun},P_{e3}^{Sun}\,|\,\theta_{12},\Delta m_{12}^2)-P_{ee}^{LE}}{\sigma_{LE}}\right)^2\\
  \chi^2_{HE}&=&\chi^2_{SK,SNO}(P_{e1},P_{e2},P_{e2}\,|\,\theta_{12},\Delta m_{12}^2)
\end{eqnarray}
we performed a statystical analysis through the following $\chi^2$ for low energy solar neutrinos:
\begin{eqnarray*}
\chi^2(P_{e1},P_{e2},P_{e2}\,|\,\theta_{12},\Delta m_{12}^2)&=&
\chi^2_{KL}(\theta_{12},\Delta m_{12}^2)+\\
&&\left(\frac{P_{ee}(P_{e1},P_{e2},P_{e2}\,|\,\theta_{12},\Delta m_{12}^2)-P_{ee}^{LE}}{\sigma_{LE}}\right)^2
\end{eqnarray*}
and a similar expression for high energy neutrinos detected during the day.

For high energy neutrinos detected during the night, we calculated:
\begin{eqnarray*}
\chi^2(P_{e1},P_{e2},P_{e2}\,|\,\theta_{12},\Delta m_{12}^2)&=&
\chi^2_{KL}(\theta_{12},\Delta m_{12}^2)+\\
&&\chi^2_{SK,SNO}(P_{e1},P_{e2},P_{e2}\,|\,\theta_{12},\Delta m_{12}^2)
\end{eqnarray*}

Marginilizing the above $\chi^2$ over the oscillation parameters, we obtain
allowed regions for probability transitions to mass eigenstates independent of
the flavour conversion mechanism operating in the Sun.

\subsubsection{Light-side (normal $12$ hierarchy)}

First we analyse what are the constrains put on $P_{ei}^{Sun}$ from Solar Neutrino data and KamLAND restricting ourselves to the light-side of the parameter space on $\theta_{12}$ ($\theta_{12}\leq \pi/4$). Although this procedure is not completely solar model independent, the only known mechanism that provides a good explanation to solar neutrino data in the dark region ($\theta_{12}>\pi/4$) involves a strong non-standard interaction, which is disfavoured~\cite{Esteban:2018ppq} by recent results on neutrino coherence scattering~\cite{Akimov:2017ade}. Anyway, on the sequence we repeat our analysis for the dark region for completeness.

\subsubsection{Low Energy}

For low-energy solar neutrinos we can see immediately that the equipartition of
mass eigenstates, with $P_{ei}=1/3$ for all $i$, is exluded, since it would
lead to $P_{ee}=1/3$. The usual MSW-LMA predicts that the conversion
probabilities solution in Sun for lower energy neutrinos are given by
their vacuum expressions:
\begin{eqnarray*}
  P_{e1}^{Sun}&=&c^2_{13}c^2_{12} \\
  P_{e2}^{Sun}&=&c^2_{13}s^2_{12} \\
  P_{e3}^{Sun}&=&s^2_{13}
\end{eqnarray*}
leading to:
\[
P_{ee}=c^4_{13}(c^4_{12}+s^4_{12})+s^4_{13}=
c^4_{13}(1-0.5\sin^2(2\theta_{12}))+s^4_{13}
\]
%Replacing $s^2_{13}=0.023$, given by the central value in~\cite{}, we have:
%\[
%\sin^2(2\theta_{12})=0.876
%\]
%which is inside KamLAND allowed region.

\vspace{1.2cm}
\begin{figure}[ht]
    \centering
    \includegraphics[width=10cm]{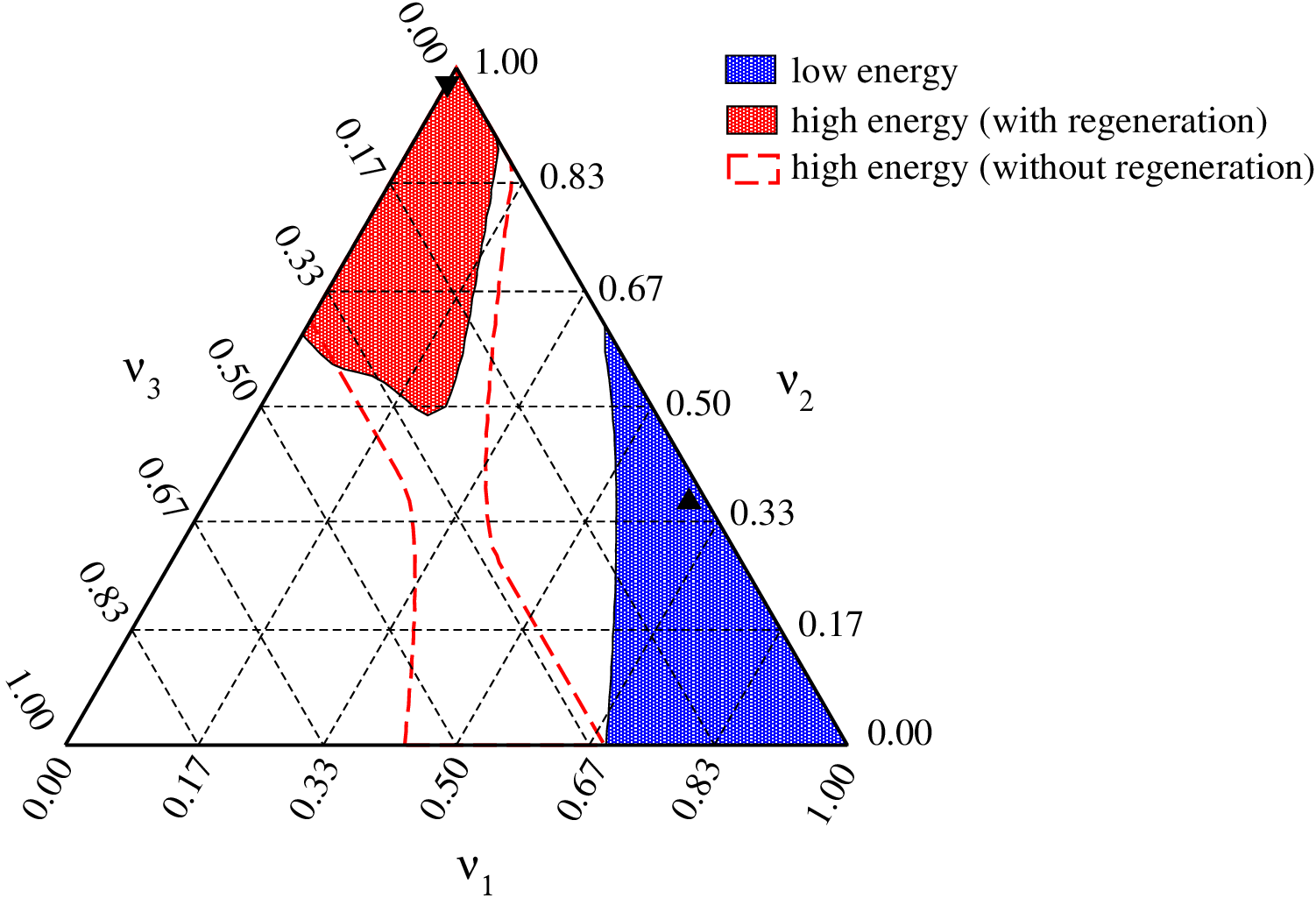}
    \caption{The allowed region for the conversion probabilities $P_{ei}$ for low-energy solar neutrinos (blue) and high-energy solar neutrinos (read). In dashed read we exclude the data from day-night asymmetry from the analysis. We restricted the analysis for the region $\theta_{12}\leq \pi/4$.
    }
    \label{fig:pei_ls}
\end{figure}

In fig.~\ref{fig:pei_ls} we can see in blue the allowed region for the values
for the conversion probabilities $P_{ei}$ given by the low-energy solar neutrino
experiments. The MSW-LMA prediction is marked with an upper triangle.

\subsubsection{High Energy}
For high-energy neutrinos the Earth matter induces changes in the probabilities,
and we have:

\begin{eqnarray*}
  P_{ee}&=&P_{e1}^{Sun} P_{1e}^{Earth}+P_{e2}^{Sun} P_{2e}^{Earth}+P_{e3}^{Sun} P_{3e}^{Earth}
\end{eqnarray*}
where $P_{ie}^{Earth}$ depends on both the oscillation parameters and the neutrino
energy. So even if the probabilities $P_{ei}$ are constant, we obtain an energy
dependent probability for neutrinos arriving at the detector during the night.
For neutrinos arriving during the day, we have the similar expression for
low-energy neutrinos:
\begin{eqnarray*}
  P_{ee}&=&c^2_{13}(P_{e1}^{Sun}c^2_{12}+P_{e2}^{Sun}s^2_{12})+s^2_{13}P_{e3}^{Sun} =0.324\pm 0.014
\end{eqnarray*}
Unlike low energy neutrinos, an equipartition of probabilities is possible,
since $P_{ee}=1/3$ is inside $1\sigma$ range. In fig~\ref{fig:pei_ls} we
present the allowed regions for solar probabilities of day high-energy neutrinos in dashed red.

When we include the night high-energy data, the regeneration effects exclude a
a large portion of allowed region. In particular, equipartition predicts a vanishing regeneration effect, disfavoured by experimental data, as can be seen in the filled red region in fig~\ref{fig:pei_ls}.

If we rewrite the survival probability during the day assuming that
$P_{e3}^{Earth}=\sin^2\theta_{13}$, which is a very good assumption for the
oscillation parameters stablished by KamLAND, we have for the survival probability during the night:

\begin{eqnarray*}
P_{ee}^N&=&P_{e1}^{Sun} P_{1e}^{Earth}+P_{e2}^{Sun} P_{2e}^{Earth}+P_{e3}^{Sun}\sin^2\theta_{13}
\end{eqnarray*}
Also, we have that $P_{2e}^{Earth}$ is always larger than $U_{2e}=\cos^2\theta_{13}\sin^2\theta_{12}$, the asymptotic value when the matter interactions is negligble. Then, to have a negative day-night asymmetry, we will need:
\[
P_{e2}^{Sun} > P_{e1}^{Sun}
\]
then excluding the portion of fig~\ref{fig:pei_ls} that splits the triangle from the lower left corner to the middle of the right edge.

Performing a $\chi^2$ analysis with the full data sample for the high-energy experiments Super-Kamiokande and SNO, using the same $\chi^2$ used in section~\ref{sec:highenergy}, we present in fig.~\ref{fig:pei_ls} in red the allowed region for high-energy neutrinos.

The MSW-LMA mechanism predicts that the high energy neutrinos, which fully felt
the resonance, leaves the Sun as a pure $\nu_2$ state (with a small fraction
of $\nu_3$ due to non-vanishing $\theta_{13}$), and is depicted as a lower triangle.

\subsection{Dark side}

\begin{figure}[ht]
    \centering
    \includegraphics[width=10cm]{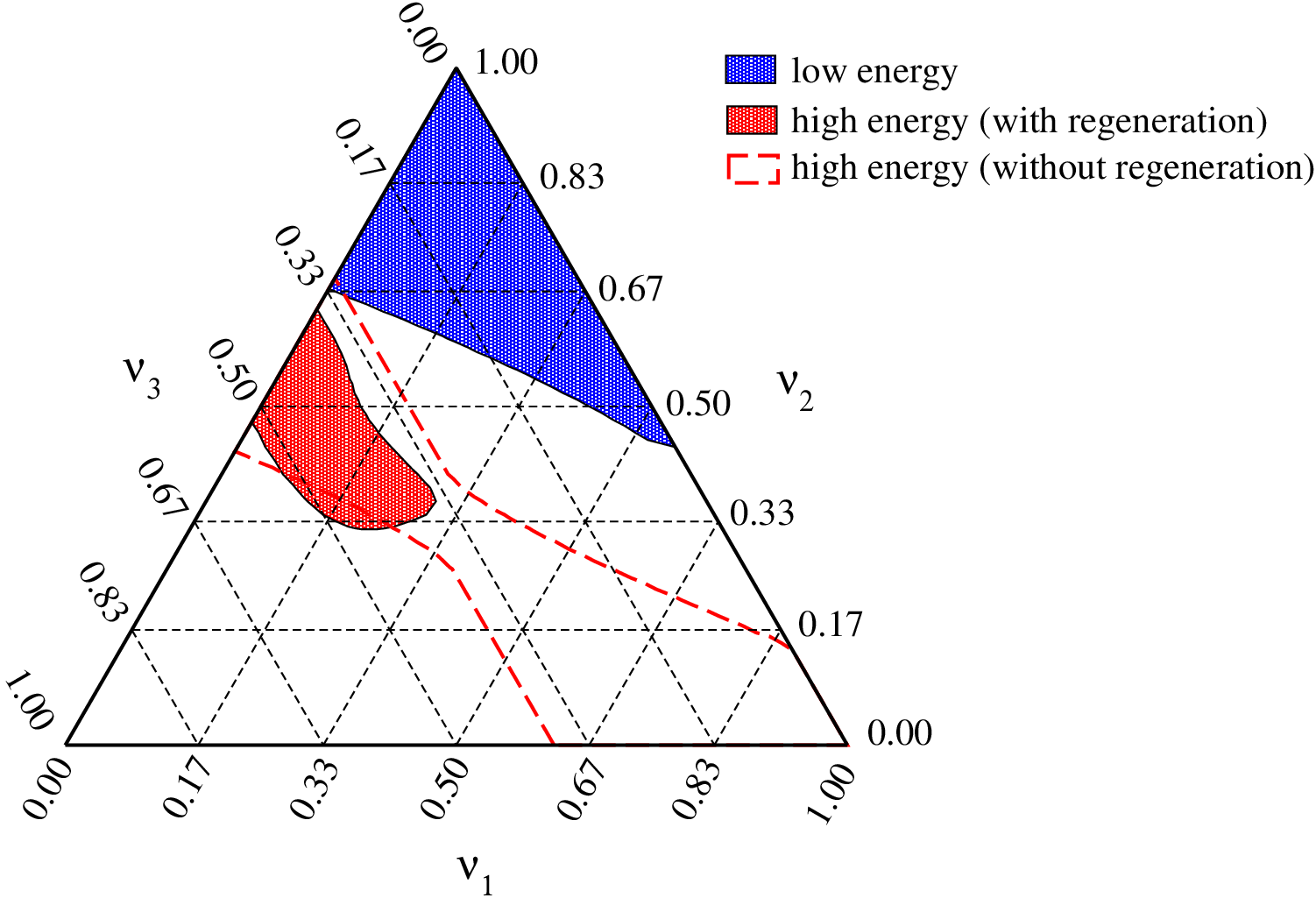}
    \caption{Same as fig.~\ref{fig:pei_ls}
      but for the dark region, $\theta_{12}>\pi/2$.}
    \label{fig:pei_ds}
\end{figure}

In fig.~(\ref{fig:pei_ds}) we present the results on the Dark-side of neutrino parameter space. If it was not for the regeneration effect, the two situations were equivalente to an exchange between $\nu_1$ and $\nu_2$. With regeneration effect, the equivalence is broken, and we get the allowed regions depicted in~\ref{fig:pei_ds}.

%\vspace{1.2cm}

In the next section we present how these informations can be used to arrive at a first constraint on new physics acting on solar neutrinos.

\section{Decoherence and Relaxation effects on Probabilities}\label{sec:decoprob}

Having constraints on the incoherente admixture of mass eignestates that composes the solar neutrino flux, it is possible to set a limit on any conversion mechanisms that alter such admixture. As can be found in several other references (for an incomplete list, see~\cite{robertot13,robertosun,balikamland,lisi,roberto,coelhoprl,robertodune,coelhopeak,robertominos,Carpio:2017nui,Coloma:2018idr}), if neutrinos are considered as a
subsystem that interacts with an environment in the Open Quantum System framework, possible modifications on neutrino conversion probabilities can be induced. As we focus in conversion between mass eingenstates, we choosed to present in a succint way the main steps for arriving at such probabilities. More details on the the procedure to obtain the conversion probabilities can be found on Appendix.

Starting with Lindblad generator, the density matrix dynamics can be expressed through the following evolution equation:
\begin{equation}
\frac{d\rho(t)}{dt}=-i[H,\rho(t)]+D(\rho)~~~,
\label{eq:rhoevol}
\end{equation}
where $\rho(t)$ is the usual density matrix, $H$ is the Hamiltonian and 
\begin{equation}
D(\rho)=+\frac{1}{2}\sum_{p=1}^{N^2-1}\left(\left[V_p,\rho(t) V_p^\dagger\right]+
\left[V_p\rho(t), V_p^\dagger\right]\right)~~~,
\label{eq:drho}
\end{equation}
with $V$ being a hermetian matrix to ensure that the system's entropy increases with time.

To solve the evolution equation it is convenient to expande it using $SU(3)$ generators and rewrite eq.~(\ref{eq:rhoevol}) in terms of the expansion coeficients. Most of the interesting phenomenological consequences of decoherence effects can be found with a diagonal $D(\rho)$ on such expansion, a choice that also guarantees complete positivity. After some algebra, we find that the matrix $D$, with the mentioned simplifying choices, can be fully parameterized as:
\begin{eqnarray}
D_{11}&=&\Gamma_1+\Gamma_8/3+\gamma_{12} \nonumber\\
D_{22}&=&\Gamma_2+\Gamma_8/3+\gamma_{12} \nonumber\\
D_{33}&=&\Gamma_3+\Gamma_8/3 \nonumber\\
D_{44}&=&\Gamma_3/4+\Gamma_4 +\gamma_{13}\nonumber\\
D_{55}&=&\Gamma_3/4+\Gamma_5 +\gamma_{13}\nonumber\\
D_{66}&=&\Gamma_3/4+\Gamma_6 +\gamma_{23}\nonumber\\
D_{77}&=&\Gamma_3/4+\Gamma_7 +\gamma_{23}\nonumber\\
D_{88}&=&\Gamma_{8}
\label{eq:gammas}
\end{eqnarray}
where the $\Gamma$'s are the parameters that induce relaxation effects, and must obey the following constraints:
\begin{eqnarray*}
\Gamma_3&=&\Gamma_1+\Gamma_2 \\
\Gamma_8&=&\frac{1}{2}\left(\Gamma_4+\Gamma_5+\Gamma_6+\Gamma_7\right)
\end{eqnarray*}
while the $\gamma$'s induce decoherence effects, and must obey the following constraints:
\[
\sqrt{2\gamma_{13}}+\sqrt{2\gamma_{23}}=\sqrt{2\gamma_{12}}
\]
or any interchange between the $\gamma$'s.

In this work we focus in the solar neutrino flux partition on mass eigenstates, so it is convenient to calculate the mass eigenstates conversion probabilities. After some algebra (shown in appendix), we obtain that such probabilities depend only on the independent parameters $\Gamma_3$ and $\Gamma_8$, through:
\begin{eqnarray}
P_{11}=P_{22}&=&\frac{1}{3}+\frac{1}{2}\exp\left[-\left(\Gamma_3+\frac{\Gamma_8}{3}\right)t\right]+\frac{1}{6}\exp[-\Gamma_8t] \nonumber\\
P_{12}=P_{21}&=&\frac{1}{3}-\frac{1}{2}\exp\left[-\left(\Gamma_3+\frac{\Gamma_8}{3}\right)t\right]+\frac{1}{6}\exp[-\Gamma_8t] \nonumber\\
P_{13}=P_{23}=P_{31}=P_{32}&=&\frac{1}{3}-\frac{1}{3}\exp[-\Gamma_8t] \nonumber\\
P_{33}&=&\frac{1}{3}+\frac{2}{3}\exp[-\Gamma_8t]
\label{eq:gammaprobabilities}
\end{eqnarray}

As it expected, when $\Gamma_{3,8}\rightarrow 0$, no transition between mass eingenstates are induced, and the system evolves adiabatically. In the other extreme, for very large values of $\Gamma_3$ and $\Gamma_8$, $P_{ij}\rightarrow 1/3$ for any $i,j$, and a complete equipartition on mass eigenstates is produced independent on the initial content of solar neutrino flux.

\section{Results}\label{sec:results}
In this section I will present the limits obtained on the new physics parameters, first with no assumption on the flavour physics at play inside the Sun, and then through a complete analysis of solar neutrino data assuming that the LMA-MSW flavour conversion mechanism as the leading effect acting on the neutrinos.

\subsection{No assumption on flavour conversion}

As discussed in Sec.~\ref{sec:nuproblimits}, fig.~\ref{fig:pei_ls} stablishes the limits on the mass eigenstates distribution on solar neutrino flux without any assumption on the neutrino flavour conversion mechanism inside the Sun. It is then straightforward to convert these constraints on limits in any new mechanism that induces conversion between mass eigenstates. For instance, since a perfect equipartition is excluded to more than 3$\sigma$, such exclusion can be converted into limits on the parameters $\Gamma_3$ and $\Gamma_8$.

\begin{figure}[ht]
    \centering
    \includegraphics[width=10cm]{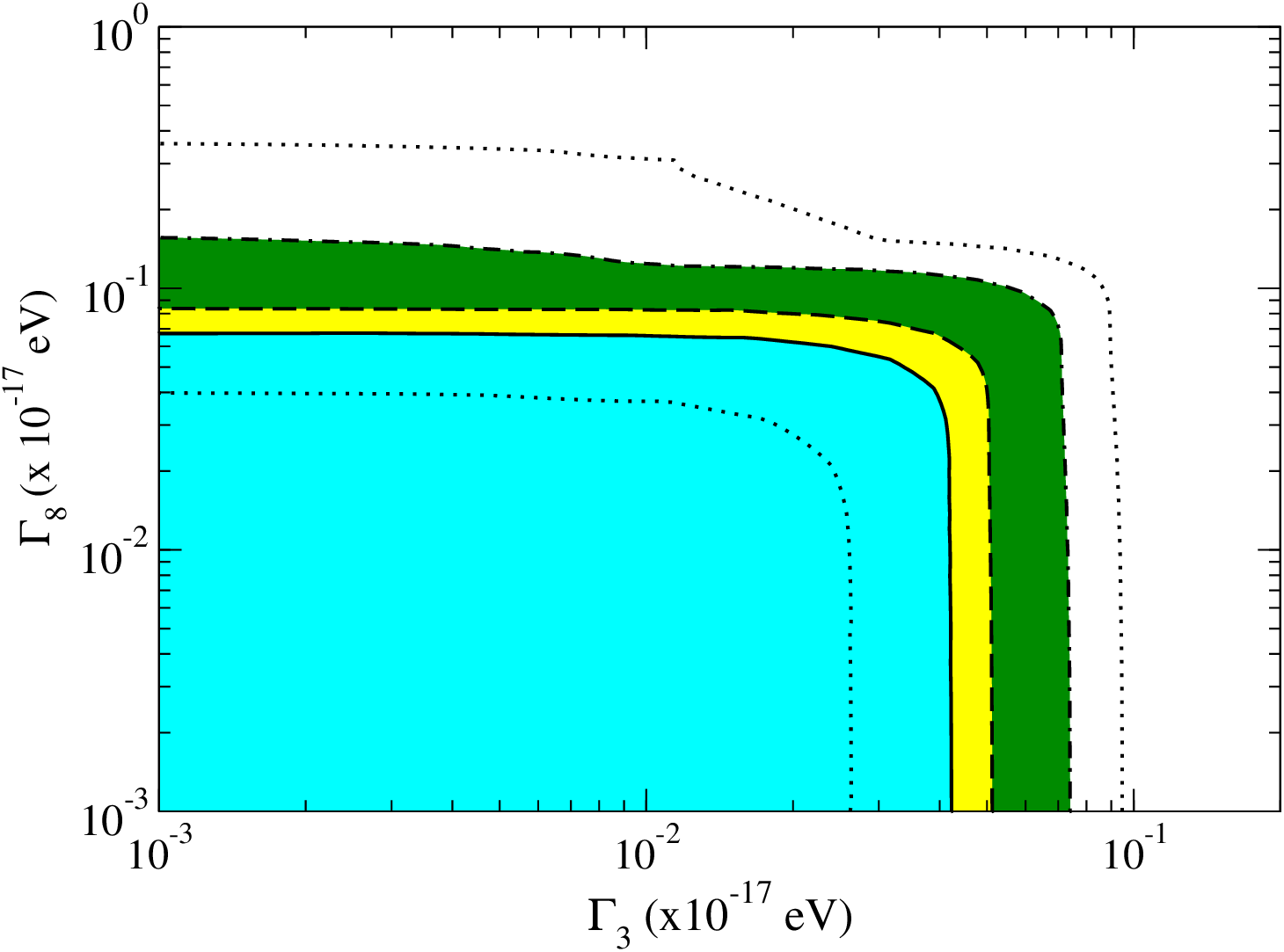}
    \caption{Allowed region on parameters $\Gamma_3$ and $\Gamma_8$ with no assumption on flavour conversion physics inside the Sun. The different regions correspond to 1$\sigma$ (dotted), 90\% C.L. (straight line), 95\% C.L. (long-dashed), 99\% C.L. (dot-dashed) and 3$\sigma$ (dotted).}
    \label{fig:chi2solarind}
\end{figure}

Introducing the probabilities presented in Eq.(\ref{eq:gammaprobabilities}) into Eq.(\ref{eq:peipie}) and performing the same statistical analysis used in Sec.2 with two extra parameters, we obtain, after minimizing over all the remaining parameters, an allowed region for the OQS parameters. We present such region at fig.~\ref{fig:chi2solarind}.
\vspace{1.2cm}

Presenting the limits one at a time, it is possible to stablish, without any assumptions on flavour conversion of solar neutrinos inside the Sun, the following limits at 1$\sigma$ (3$\sigma$):
\begin{equation}
  \Gamma_3< 1.7\,(7.2)\times 10^{-19}~{\rm eV} ~~~;~~~
  \Gamma_8< 2.2\,(15.0)\times 10^{-19}~{\rm eV}~ .
\end{equation}

\subsection{Complete statistical analysis}

The exercise presented in previous sections consists of a very robust way to limit new physics on neutrino sector from neutrino flux produced in astrophysics objects, where it is expected that the neutrinos arrive at Earth as an incoherent admixture of mass eigenstates, and the uncertainties on the initial neutrino flux are large. For solar neutrinos, only the first of these statements applies, so we can proceed a further step and include our knowledge on flavour conversion inside the Sun to stablish a stronger bound on the new physics analysed here. We present in this section the results of such analysis.

For the solar neutrino data, we perform a statistical analysis with the full spectral data from Super-Kamiokande phases I, III and IV~\cite{Hosaka:2005um,Fukuda:2002pe,Abe:2016nxk}, the combined analysis of all three SNO phases~\cite{Aharmim:2011vm}, Borexino~\cite{Agostini:2018uly}, combined Gallex+GNO~\cite{Kaether:2010ag}, SAGE~\cite{Abdurashitov:2009tn} and Homestake~\cite{Cleveland:1998nv}. For KamLAND data we used their 2008 spectral data~\cite{Abe:2008aa}.

The statistical procedure follows the same one presented in previous articles~\cite{deHolanda:2008nn,deHolanda:2003nj}, with two free parameters related to standard neutrino parameters, $\Delta m^2_{21}$ and $\theta_{12}$, and two new free parameters related to new physics, $\Gamma_3$ and $\Gamma_8$. The mixing angle $\theta_{13}$ is fixed at the best fit point of a global analysis~\cite{Esteban:2018azc}.

\begin{figure}[ht]
    \centering
    \includegraphics[width=10cm]{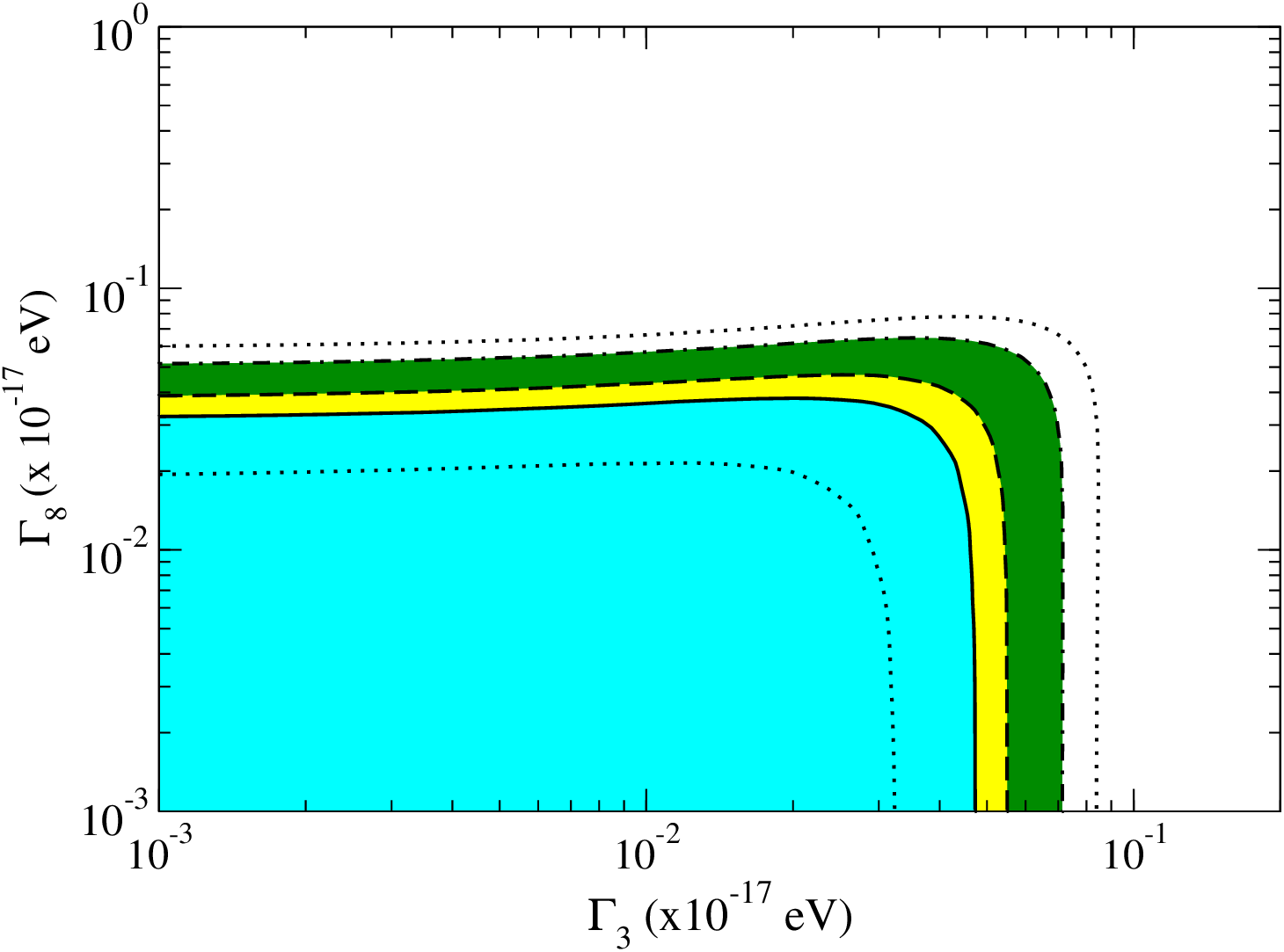}
    \caption{Allowed region on parameters $\Gamma_3$ and $\Gamma_8$ using the full knowledge of MSW effect inside the Sun. The different regions correspond to 1$\sigma$ (dotted), 90\% C.L. (straight line), 95\% C.L. (long-dashed), 99\% C.L. (dot-dashed) and 3$\sigma$ (dotted).}
    \label{fig:chi2all}
\end{figure}

After minimizing over the standard neutrino parameters, we obtain the allowed region shown in  fig.~\ref{fig:chi2all}. We obtain a slight enhancement on the limits in the non-standard parameters when we use the full knowledge of the neutrino evolution inside the Sun.

\vspace{1.2cm}

Presenting the limits one at a time, the solar neutrinos + KamLAND data provide the following limits at 1 $\sigma$ (3 $\sigma$):
\begin{equation}
  \Gamma_3< 1.1\,(6.5)\times 10^{-19} {\rm eV} ~~~;~~~ \Gamma_8< 2.1\,(7.1)\times 10^{-19} {\rm eV}~ .
\end{equation}

\section{Conclusions}\label{sec:conclusions}

We presented in this work a novel procedure to analyse the neutrino flux coming from astrophysical objects, focusing on the mass eigenstates partition, since it is expected that the neutrino flux arrives as an incoherent admixture of such eigenstates. Using solar neutrino flux data, we infered the limits on relaxation parameters in a Open Quantum System formalism with no assumptions on the neutrino flavour physics inside the Sun. When compared with limits obtained with full knowledge of solar neutrino production and MSW flavour conversion mechanism, the limits could be only slightly enhanced, demonstrating that the analysis with no assumptions on flavour conversion mechanism inside the Sun is quite robust. We pretend to use this kind of analysis in other astrophysical objects, such as supernovae and high-energy neutrino sources, where the uncertainty of neutrino production and flavour conversion mechanisms are uncertain.

\section*{Acknowledgements}

 Work partially supported by CNPq (grant no. 310952/2018-2) and FAPESP (grant no. 2014/19164-6 and no. 2019/15600-0). The author would like to thank D.~R.~Gratieri and R.~L.~N.~Oliveira, with whom discussions on the early stage of the work were fundamental to its production.

\bibliography{bibliography}{}
\bibliographystyle{ieeetr}

\appendix
\section{OQF Formalism}

It is possible to write the decoherence operator in~eq.(\ref{eq:drho}) as:
\[
D(\rho)=
+\frac{1}{2}\sum_{p=1}^{N^2-1}\left[\left[V_p,\rho(t)\right], V_p\right]
\]

Expanding using Gell-Mann matrices:
\[
V_p=v_{pi}\lambda_i~~~;~~~ \rho=\rho_j\lambda_j
\]
we have:
\[
D(\rho)=
+\frac{1}{2}\sum_{p,i,j,k}v_{pi}v_{pk}\rho_j[[\lambda_i,\lambda_j],\lambda_k]
\]
and using the SU(3) structure constants:
\[
\left[\frac{\lambda_i}{2},\frac{\lambda_j}{2}\right]=
f_{ijk}\frac{\lambda_k}{2}
\]
we have (with sumation implicit):
\[
D(\rho)=
+v_{pi}v_{pk}\rho_jf_{ijl}[\lambda_l,\lambda_k]=
+2v_{pi}v_{pk}\rho_jf_{ijl}f_{lkm}\lambda_m
\]
Finally, writing the evolution equation in terms of a system on $\rho_i$'s:
\[
\frac{d\rho_l}{dt}=(...)+D_{jm}\rho_j
\]
where 
\[
D_{jm}=2v_{pi}v_{pk}f_{ijl}f_{kml}
\]

Replacing the SU(3) structure constants:
\[
f^{123}=1~~;~~f^{147}=f^{165}=f^{246}=f^{257}=f^{345}=f^{376}=\frac{1}{2} ~~;~~
 f^{458}=f^{678}=\frac{\sqrt{3}}{2}
\]
we have for the diagonal entries:
\[
D_{11}=2\sum_{p,i,l}(v_{pi})^2(f_{i1l})^2
\]
Replacing all non-null entries:
\[
D_{11}=2\sum_p\left((v_{p2})^2+(v_{p3})^2+
\frac{1}{4}(v_{p4})^2+\frac{1}{4}(v_{p5})^2+
\frac{1}{4}(v_{p6})^2+\frac{1}{4}(v_{p7})^2\right)
\]
Renaming:
\[
\vec{a}_i=\left(\{v_{pi}\},p=1,8\right)
\]
we obtain:
\[
D_{11}=2a_2^2+2a_3^2+\frac{1}{2}a_4^2+\frac{1}{2}a_5^2+\frac{1}{2}a_6^2+\frac{1}{2}a_7^2
\]
and, similarly:
\[
D_{22}=2a_1^2+2a_3^2+\frac{1}{2}a_4^2+\frac{1}{2}a_5^2+\frac{1}{2}a_6^2+\frac{1}{2}a_7^2
\]
\[
D_{33}=2a_1^2+2a_2^2+\frac{1}{2}a_4^2+\frac{1}{2}a_5^2+\frac{1}{2}a_6^2+\frac{1}{2}a_7^2
\]
\[
D_{88}=\frac{3}{2}\left(a_4^2+a_5^2+a_6^2+a_7^2\right)
\]
For the $44$ element:
\[
D_{44}=2v_{pi}v_{pk}f_{i4l}f_{k4l}
\]
we have (renaming $\sqrt{3}a_8\rightarrow a_8$):
\begin{eqnarray*}
D_{44}&=&\frac{1}{2}\left(a_1^2+a_2^2+a_3^2+a_5^2+a_6^2+a_7^2\right)+
\frac{3}{2}\left(a_5^2+a_8^2\right)+
4v_{p3}v_{p8}\frac{\sqrt{3}}{4}\\
&=&\frac{1}{2}\left(a_1^2+a_2^2+4a_5^2+a_6^2+a_7^2\right)+
\frac{1}{2}\left(\vec{a}_3+\vec{a}_8\right)^2
\end{eqnarray*}
and, similarly:
\[
D_{55}=\frac{1}{2}\left(a_1^2+a_2^2+4a_4^2+a_6^2+a_7^2\right)+
\frac{1}{2}\left(\vec{a}_3+\vec{a}_8\right)^2
\]
\[
D_{66}=\frac{1}{2}\left(a_1^2+a_2^2+a_4^2+a_5^2+4a_7^2\right)+
\frac{1}{2}\left(\vec{a}_3-\vec{a}_8\right)^2
\]
\[
D_{77}=\frac{1}{2}\left(a_1^2+a_2^2+a_4^2+a_5^2+4a_6^2\right)+
\frac{1}{2}\left(\vec{a}_3-\vec{a}_8\right)^2
\]
Having only decoherence, and no relaxation, it is possible to show that:
\[
a_1=a_2=a_4=a_5=a_6=a_7=0
\]
and we renamed the non-vanishing parameters as:
\begin{eqnarray*}
  D_{11}=D_{22}&=&2a_3^2=\gamma_{12} \\
  D_{44}=D_{55}&=&\frac{1}{2}\left(\vec{a}_3+\vec{a}_8\right)^2=\gamma_{13} \\
  D_{66}=D_{77}&=&\frac{1}{2}\left(\vec{a}_3-\vec{a}_8\right)^2=\gamma_{23} \\
\end{eqnarray*}
For colinear $\vec{a}_3$ and $\vec{a}_8$ the following constraint are obtained:
\[
\sqrt{2\gamma_{13}}+\sqrt{2\gamma_{23}}=\sqrt{2\gamma_{12}}
\]
or any interchange between the $\gamma$'s.

Introducing relaxation, a convenient nomenclature would be to set:
\begin{eqnarray*}
\Gamma_1&=&2a_2^2 \\
\Gamma_2&=&2a_1^2 \\
\Gamma_3&=&2a_1^2+2a_2^2 \\
\Gamma_4&=&2a_5^2+\frac{1}{2}(a_6^2+a_7^2) \\
\Gamma_5&=&2a_4^2+\frac{1}{2}(a_6^2+a_7^2) \\
\Gamma_6&=&\frac{1}{2}(a_4^2+a_5^2)+2a_7^2 \\
\Gamma_7&=&\frac{1}{2}(a_4^2+a_5^2)+2a_6^2 \\
\Gamma_8&=&\frac{3}{2}\left(a_4^2+a_5^2+a_6^2+a_7^2\right)
\end{eqnarray*}
and with this choice we obtain:
\begin{eqnarray*}
D_{11}&=&\Gamma_1+\Gamma_8/3+\gamma_{12} \\
D_{22}&=&\Gamma_2+\Gamma_8/3+\gamma_{12} \\
D_{33}&=&\Gamma_3+\Gamma_8/3 \\
D_{44}&=&\Gamma_3/4+\Gamma_4 +\gamma_{13}\\
D_{55}&=&\Gamma_3/4+\Gamma_5 +\gamma_{13}\\
D_{66}&=&\Gamma_3/4+\Gamma_6 +\gamma_{23}\\
D_{77}&=&\Gamma_3/4+\Gamma_7 +\gamma_{23}\\
D_{88}&=&\Gamma_{8}
\end{eqnarray*}
where the $\Gamma$'s must obey the following constraints:
\begin{eqnarray*}
\Gamma_3&=&\Gamma_1+\Gamma_2 \\
\Gamma_8&=&\frac{1}{2}\left(\Gamma_4+\Gamma_5+\Gamma_6+\Gamma_7\right)
\end{eqnarray*}

To calculate survival probabilities, we use:
\[
P_{\alpha\beta}=\sum_i2\rho_{i,\alpha}(0)\rho_{i,\beta}(t)
\]
and to focus on conversion probabilities between mass eigenstates, we would
use:
\begin{eqnarray*}
  \nu_1&:& \rho_0=\frac{1}{\sqrt{6}}~~~;~~~\rho_3=+\frac{1}{2}~~~;~~~\rho_8=+\frac{1}{2\sqrt{3}} \\
  \nu_2&:& \rho_0=\frac{1}{\sqrt{6}}~~~;~~~\rho_3=-\frac{1}{2}~~~;~~~\rho_8=+\frac{1}{2\sqrt{3}} \\
  \nu_3&:& \rho_0=\frac{1}{\sqrt{6}}~~~;~~~\rho_3=0~~~;~~~\rho_8=-\frac{1}{\sqrt{3}} 
\end{eqnarray*}

Working on diagonal basis, where $H=h_3\lambda_3+h_8\lambda_8$ is easy to see
that the evolution of $\rho_3$ and $\rho_8$ are completely decoupled,
and then:
\begin{eqnarray*}
\rho_3(t)&=&\exp\left[-\left(\Gamma_3+\frac{\Gamma_8}{3}\right)t\right]\rho_3(0)\\
\rho_8(t)&=&\exp\left[-\Gamma_8t\right]\rho_8(0)
\end{eqnarray*}
which leaves to:
\begin{eqnarray*}
P_{11}=P_{22}&=&\frac{1}{3}+\frac{1}{2}\exp\left[-\left(\Gamma_3+\frac{\Gamma_8}{3}\right)t\right]+\frac{1}{6}\exp[-\Gamma_8t] \\
P_{12}=P_{21}&=&\frac{1}{3}-\frac{1}{2}\exp\left[-\left(\Gamma_3+\frac{\Gamma_8}{3}\right)t\right]+\frac{1}{6}\exp[-\Gamma_8t] \\
P_{13}=P_{23}=P_{31}=P_{32}&=&\frac{1}{3}-\frac{1}{3}\exp[-\Gamma_8t] \\
P_{33}&=&\frac{1}{3}+\frac{2}{3}\exp[-\Gamma_8t] 
\end{eqnarray*}

\end{document}